# Investigating Ionic Diffusivity in Amorphous Solid Electrolytes using Machine Learned Interatomic Potentials

Aqshat Seth, Rutvij Pankaj Kulkarni, Sai Gautam Gopalakrishnan[*]

Department of Materials Engineering, Indian Institute of Science, Bengaluru 560012, India

*Corresponding author: saigautamg@iisc.ac.in

## Abstract

Due to its immense importance as an amorphous solid electrolyte in thin-film devices, lithium phosphorous oxynitride (LiPON) has garnered significant scientific attention. However, investigating $Li^+$ transport within the LiPON framework, especially across a Li||LiPON interface, has proven challenging due to its amorphous nature and varying stoichiometry, necessitating large supercells and long timescales for computational models. Notably, machine learned interatomic potentials (MLIPs) can combine the computational speed of classical force fields with the accuracy of density functional theory (DFT), making them the ideal tool for modelling such amorphous materials. Thus, in this work, we train and validate the neural equivariant Interatomic potential (NequIP) framework on a comprehensive DFT-based dataset consisting of 13,454 chemically relevant structures to describe LiPON. With an optimized training (validation) energy and force mean absolute errors of 5.5 (6.1) meV/atom and 13.6 (13.2) meV/Å, respectively, we employ the trained potential in model Li-transport in both bulk LiPON and across a Li||LiPON interface. Amorphous LiPON structures generated by the optimized potential do resemble those generated by *ab initio* molecular dynamics, with N being incorporated on non-bridging apical and bridging sites. Subsequent analysis of $Li^+$ diffusivity in the bulk LiPON structures indicates broad agreement with computational and experimental literature so far. Further, we investigate the anisotropy in $Li^+$ transport across the Li(110)||LiPON interface, where we observe Li-transport across the interface to be one order-of-magnitude slower than Li-motion within the bulk Li and LiPON phases. Nevertheless, we note that this anisotropy of Li-transport across the interface is minor and do not expect it to cause any significant impedance buildup. Finally, our work highlights the efficiency of MLIPs in enabling high-fidelity modelling of complex non-crystalline systems over large length and time scales.

# Introduction

The utilization of solid electrolytes in lithium (Li)-based energy storage technology is an active area of research as solid electrolytes allow for the use of lithium metal anodes, thereby improving the energy density and safety of next generation Li-based batteries.[1–8] Specifically, amorphous materials being used as solid electrolytes is particularly promising, given their wide compositional stability (or flexibility) that allows for significant change in ionic content without adverse phase transformations,[9] their lack of grain boundaries that mitigates charge transfer impedance[10] and the absence of electrostatic or structural inhomogeneities[11] that can result in the nucleation of dendrites.[12] An example of an amorphous solid electrolyte is the lithium phosphorous oxynitride (of chemical formula $Li_xPO_yN_z$ where $x = 2y+3z-5$), commonly referred to as LiPON, which has been demonstrated in thin-film energy storage devices.[13] Synthesized first by Bates et al.,[14] LiPON is typically made by incorporating N into $Li_3PO_4$ via radio-frequency (RF) magnetron sputtering. The remarkable properties exhibited by LiPON are often attributed to the incorporation of N into the $Li_3PO_4$ structure, including electrochemical and mechanical stability,[15] low electronic conductivity ($10^{-15}$ to $10^{-12}$ S/cm),[16] moderate ionic conductivity ($3\times10^{-6}$ S/cm),[17–19] high critical current density ($> 10$ mA/cm$^2$)[20] and excellent cyclability against lithium metal anodes.[21,22] Despite the well-established properties of LiPON, the local structure of the electrolyte and the specific role of N in enhancing the performance of LiPON, especially on suppressing Li-dendrite formation, still remain uncertain. The wide compositional range of LiPON, besides its amorphous structure, makes both experimental and computational investigations of the system challenging.

Prior analyses of the N 1*s* X-Ray photoelectron spectroscopy (XPS) data on LiPON thin films have suggested that N atoms in LiPON crosslink by bonding to two or three phosphate tetrahedra, resulting in the formation of double-bridging ($N_d$) and triple-bridging ($N_t$) N sites, respectively.[23–28] Such crosslinking by N leads to a "mixed anion effect",[29] which can provide Li$^+$ with interconnected, low activation energy pathways, thus improving their diffusivity with respect to bulk, crystalline $Li_3PO_4$.[30] Wang et al., based on X-Ray diffraction and chromatography data along with the initial XPS data proposed the presence of apical or non-bridging N atoms ($N_a$), leading to isolated $PO_3N$ tetrahedra, besides a small amount of $N_d$.[31] Other studies have claimed LiPON resembles metaphosphate glasses with extended chains of phosphate tetrahedra linked by N or O atoms or layered structures with Li and P rich regions,[32,33] which does not match with the presence of isolated $PO_4$ tetrahedra in the precursor phase of $Li_3PO_4$.

Computational studies, employing tools such as density functional theory (DFT) based calculations, ab-initio molecular dynamics (AIMD), and classical molecular dynamics (MD) based on machine learned interatomic potentials (MLIPs) have also been used to shed light on the structural features contributing to the enhanced electrochemical properties of LiPON. For example, Sicolo et al. observed N atoms in the form of both $N_d$ and $N_t$ in their melt-quench generated amorphous LiPON structure, albeit with a stoichiometry ($Li_{1.25}PO_2N_{0.75}$) resembling bulk phosphate glasses than the thin-film solid electrolytes.[34] Lacivita *et al.* observed N to be incorporated as $N_a$ and $N_d$ (and not $N_t$) in their AIMD-generated LiPON structures[35] and proposed that the $N_d$ atoms densified the LiPON framework leading to the destabilization of $Li^+$ and an improved Li-mobility. Further, the authors found the $N_a$:$N_d$ ratio increased as Li content in LiPON increased, eventually resulting in all N occupying only $N_a$ sites at a composition of $Li_{3.38}PO_{3.62}N_{0.38}$. Subsequently, the authors proposed alternative assignments of existing experimental data that fully avoid assigning N to $N_t$ sites, by using a combination of computational techniques with neuron scattering and infrared (IR) spectroscopy.[36] Using solid-state nuclear magnetic spectroscopy (ssNMR) with AIMD, Marple et al. explored the short-range environment of the phosphorous atoms in LiPON and reported four distinct phosphate tetrahedra configurations, namely $PO_4^{3-}$, $PO_3N^{4-}$ ($N_a$), $P_2O_6N^{5-}$ ($N_d$) and $P_2O_7^{4-}$.[37] Nevertheless, the Li-transport across a Li||LiPON interface has not been probed in detail so far.

Given that amorphous systems often require large length and long time scales to sample the system dynamics well, classical MD powered by MLIPs are highly pertinent to model amorphous systems, given their high accuracy and low computational costs compared to DFT/AIMD.[38–41] Typically, MLIPs are trained on a DFT-computed smaller supercell dataset to mathematically approximate the potential energy surface (PES) of several configurations that a system can exhibit.[42] Among the diverse set of MLIPs available, graph neural network (GNN)-based potentials, where the atomic structures are mapped onto graph(s) containing nodes (atoms) and edges (bonds) that are subsequently convoluted to incorporate short- and long-range interactions and invariant/equivariant symmetry constraints, have emerged as highly data-efficient architectures for deep learning PES of materials.[43–45] The neural equivariant interatomic potential (NequIP)[46] is particularly promising among GNN-based potentials, since NequIP utilizes higher-order equivariant tensors that preserve translational, rotational, and permutational invariance,[45] allowing it to build "flexible" potentials with high accuracy and fewer training structures compared to other MLIPs.

In this study, we develop a DFT and AIMD computed LiPON-based dataset and use it to train and optimize NequIP to model $Li^+$ transport within the bulk amorphous framework and across a Li||LiPON interface. We generate over 13,000 configurations for training and validating the NequIP model, via DFT calculations on LiPON precursors, strained bulk structures, lattices with varying Li concentrations, slab/surface configurations, and AIMD simulations. Subsequently, we use the trained NequIP model to generate amorphous structures and probe $Li^+$ diffusivity using MD simulations. Importantly, we find the predicted amorphous LiPON structures to resemble structures presented in prior AIMD-based simulations, highlighting NequIP's accuracy in predicting structural features of LiPON. To investigate $Li^+$ transport in LiPON based thin-film devices, we simulate $Li^+$ diffusion through bulk amorphous LiPON as well as across a Li(110)||LiPON interface. Our results indicate $Li^+$ diffusivity to be anisotropic across the Li||LiPON interface, varying by roughly one order of magnitude

within the bulk and across the interface, with the degree of anisotropy dependent on the simulation temperature. We hope our study and the potentials we have constructed allow for further exploration of bulk LiPON and Li||LiPON interfaces and facilitate the utilization of MLIPs in the investigation of other amorphous solid electrolytes.

## Methods

### Dataset Generation

Our dataset can be broadly divided into five categories of structures, which were generated using different procedures, namely, *i*) strained, *ii*) Li-rich, *iii*) Li-poor, *iv*) melt-quench, and *v*) slab-based. Using an initial set of 19 different chemical systems (as listed in **Table S1** of the Supporting Information – SI), including elemental Li, $Li_3P$, $Li_3N$, $LiPN_2$, and $Li_3PO_4$, we used the pymatgen[47] package to generate various strained configurations which can capture the local environment while $Li^+$ diffuses within the amorphous LiPON framework. We induced hydrostatic (-9% to +10%), monoclinic (-8.65% to +8.65%), and orthorhombic (-10% to +10%) strains on the DFT-relaxed bulk structures, amounting to a total of 886 strained configurations. To capture the variations in Li content, we generated 256 defective compositions, i.e., by replacing one O with a N in the unit cell as well as in a 3×2×2 supercell of $Li_3PO_4$. To balance the charge within the defective structures (i.e., one O replaced with a N), we enumerated symmetrically distinct ways to add a single lithium mimicking Li-rich conditions, and also enumerated ways to remove both an O and a Li to model Li-poor concentrations. Additionally, we constructed 74 Li-rich structures by enumerating the removal of 1 P and the concomitant addition of 5 Li within $Li_7PN_4$ supercells. To eventually model local environments that may be encountered in a Li||LiPON interface, we incorporated 1,219 slab-based configurations from several chemistries, including elemental Li, $Li_3P$, $Li_2O$, and $Li_3N$. The slabs were generated with a vacuum spacing of 20 Å and different slab thickness of 10 Å to 30 Å. Finally, to capture the amorphous nature of LiPON during training, we generated 11,000 amorphous structures using the melt and quench approach in AIMD simulations, resulting in a total dataset of 13,454 configurations. We have made available the entire dataset used for training and validating our NequIP model, as part of our GitHub repository (see Data availability section).

### Training and validation

NequIP expresses the PES of a given structure through the summation of atomic energies, which in turn are functions of their corresponding local environment, while atomic forces are obtained as the gradient of the total potential energy. NequIP makes use of equivariant convolutions in 3D Euclidean space (or E(3)-equivariant)[48] for modelling the interactions among its geometric-tensors-based features. The convolution filter in NequIP is a product of equivariant and learnable radial functions and spherical harmonic functions.[49] We split our overall 13,454 structure dataset randomly in a 90:10 ratio for training and validation, respectively, where we optimized the hyperparameters to obtain the least mean absolute errors (MAEs) on atomic forces in the validation set. To investigate the effect of equivariance on the accuracy of the trained potential, we trained a NequIP model with the optimized hyperparameters but with the feature set restricted to only consist of scalars. The final set of optimized hyperparameters is in **Table S2** along with the final energy and force MAEs that we obtained on our train and validation sets. After constructing the optimized NequIP model, we performed melt-quench simulations using this model to generate amorphous LiPON structures, and subsequently calculate Li$^+$ diffusivity in bulk LiPON and across the Li||LiPON interface.

**DFT and AIMD calculations**

We used the Vienna ab initio simulation package (VASP)[50,51] with the projector augmented wave (PAW)[52,53] potentials to generate the entire DFT-calculated dataset. For treating the electronic exchange and correlation, we used the Perdew-Burke-Ernzerhof (PBE) functionalization of the generalized gradient approximation (GGA).[52] We fixed the kinetic energy cutoff to 520 eV and sampled the Brillouin zone using Γ-centered Monkhorst−Pack *k*-point meshes of density 32/Å (i.e., a minimum of 32 *k*-points were sampled across a reciprocal space lattice vector of 1 Å$^{-1}$) for relaxing all structures. For all structures, we relaxed the cell volume, cell shape, and ionic positions without symmetry constraints until the total energies and atomic forces converged within $10^{-5}$ eV and |0.03| eV/Å, respectively. We performed only a single self-consistent-field (SCF) calculation for all strained configurations (till total energies converged within $10^{-5}$ eV). Additionally, we performed a single SCF calculation on the relaxed geometries of all non-AIMD structures at an energy cut-off of 400 eV, to ensure that our energy scales are comparable with the AIMD simulations (see below).

To generate the DFT-calculated amorphous training dataset, we melted the calculated ground-state structures for Li$_3$PO$_4$ and Li-rich defective structures by rapid heating till 3000 K, using AIMD simulations. Subsequently, we quenched the molten structures from 3000 K to approximately 0 K, at a rate of 250 K/ps. For Li$_3$P, Li$_2$O, and Li$_3$N we followed a similar melt-quench approach where we heated the corresponding relaxed unit cells up to 2000 K, 2000 K, and 1000 K, respectively. Our choice of temperatures up to which we heated the systems considered were based on the corresponding melting points of the compounds, namely, 1478 K for Li$_3$PO$_4$, 742 K for Li$_3$P, 1711 K for Li$_2$O and 1087 K for Li$_3$N. For all AIMD simulations, we used the NVT ensemble with the Nose-Hoover thermostat,[54–56] a time step of 2 fs, and a kinetic energy cutoff of 400 eV. We used a lower kinetic energy cut-off in our AIMD calculations to reduce computational costs.

**Generating amorphous LiPON**

We used the trained NequIP potential to generate the 'equilibrated' and amorphous LiPON structures using the large-scale atomic/molecular massively parallel simulator (LAMMPS)[57] package. The initial configuration for the LAMMPS simulations was generated by taking a $2 \times 2 \times 2$ Li$_3$PO$_4$ supercell of 128 atoms (i.e., 16 formula units), replacing five random oxygen atoms with nitrogen atoms and balancing the charges by removing Li and O atoms, leading to a final composition of Li$_{2.94}$PO$_{3.5}$N$_{0.31}$. To create the equilibrated structures at different temperature (i.e., structures not necessarily intended to become amorphous), we subjected the initial crystalline LiPON structure to NVT simulations of 100 ps, with a time step of 2 fs at temperatures of 600 K, 900 K, 1200 K and 1500 K. For generating amorphous LiPON, we started by melting the initial LiPON structure under NVT at 2000 K for 10 ps, with a timestep of 1 fs, followed by a quench from 2000 K to 250 K at a rate of 250 K/ps.

**Construction of the Li||LiPON interface**

The choice of Li(110) is motivated by its high stability (as indicated by the low calculated surface energy of 0.0309 eV/Å$^2$) and low lattice parameter mismatch between the two systems. We chose the 2×2×1 supercell of the Li(110) slab to interface with the MLIP-based melt-quench LiPON structure. Subsequently, we modified the LiPON structure such that its lattice parameters match those of the 2×2×1 Li(110) slab along the *a* and *b* directions. This transformation of the LiPON structure leads to a contraction of 1.74% in the surface area along the *a-b* plane. To offset this contraction and preserve the initial volume of the LiPON structure, we expanded the LiPON structure along the *c* direction by 1.39%. The Li(110) slab was given a thickness of 19.5 Å to distinguish the interfacial behavior from that of the bulk. The total interface structural model thickness was fixed at 36.5 Å with a gap of 2 Å between the Li(110) slab and the LiPON slab.

**Li$^+$ diffusivity**

We modelled Li$^+$ diffusivity within the amorphous LiPON structures via NVT ensemble simulations of a minimum of 100 ps with a time step of 5 fs at different temperatures, using LAMMPS powered by our NequIP model. Similarly, we modelled Li$^+$ motion across the Li||LiPON interface for 75 ps with a time step of 5 fs at different temperatures. In both amorphous and interfacial systems, ionic diffusivity ($D$) is calculated as the slope of the mean square displacement (MSD) of Li-atoms over time interval ($\Delta t$), as given by **Equation 1**, where $d$ refers to the dimensionality of the system. We measure the MSD by averaging over the MSD of each Li ion within the entire simulation, i.e., our calculated $D$ corresponds to the tracer diffusivity of Li-ions.

$$D = \frac{MSD(\Delta t)}{2d\Delta t} \qquad (1)$$

However, a simple slope of MSD versus $\Delta t$ does not account for the ballistic and vibrational motion of the ions. Hence, we followed the procedure proposed by He et al.[58] which neglects the time step increments until the MSD reaches a value of $0.5a^2$, where *a* denotes the average distance between two neighboring Li sites. Also, we ensured the linearity of MSD versus $\Delta t$ while estimating $D$. Using the calculated diffusivity, we also calculated the ionic conductivity in amorphous LiPON via the Nernst-Einstein relation, as in **Equation 2**.

$$\sigma = \frac{Nq^2}{VkT}D \qquad (2)$$

where $V$, $N$, $q$, $k$, and $T$ denote the volume of the system, the number of mobile Li ions, the charge on a Li$^+$, the Boltzmann constant, and the temperature, respectively.

# Results

## AIMD simulations

Examples of the AIMD-generated amorphous Li-rich LiPON and Li$_3$PO$_4$ structures compositions along with their respective radial distribution (RDFs) plots are shown in **Figure 1**. Note that the structures and RDFs are snapshots taken at 250 K after the systems have been quenched fully. The green and red spheres in panels a and c of **Figure 1** indicate Li and O atoms, while the purple polyhedral indicate PO$_4$ groups. N atoms in the LiPON structure are highlighted by blue spheres and the PO$_3$N tetrahedral groups that are a part of colored in blue in **Figure 1a**. In panels b and d, blue, black, green, and red lines indicate Li-Li, Li-P, Li-O, and Li-N neighbors (or bonds), respectively. Sample RDFs from melt-quench AIMD simulations of Li$_3$N, Li$_3$P, and Li$_2$O are compiled in **Figure S1**, with Li$_3$N and Li$_3$P showing distinct signatures of amorphous phases.

Importantly, the lack of uniform and periodic sharp peaks in the RDFs plots (panels b and d of **Figure 1**) indicate that the structures are completely disordered after the melt-quench process within our AIMD simulations. Also, the broad peaks of all types of Li-based neighbors beyond 4 Å in both LiPON and Li$_3$PO$_4$ indicate the lack of any long-range order, a typical signature observed in amorphous structures. Comparing the two calculated RDFs, we notice that the distribution of Li-O bonds remains largely unaffected even after the incorporation of N in the amorphous-Li$_3$PO$_4$, while Li-Li bonds show minor variations. We do observe a significant change in the Li-P profiles of LiPON and Li$_3$PO$_4$, with the two distinct peaks in Li$_3$PO$_4$ (at 2.51 and 3.14 Å, see **Figure 1d**) being replaced by a single, broad peak at ~3 Å in LiPON (see **Figure 1b**), suggesting a change in distribution of the Li atoms around the P-based tetrahedral groups in amorphous LiPON. Notably, we observe the distribution of the Li-N bonds in LiPON to closely follow that of Li-O bonds (in LiPON/Li$_3$PO$_4$), especially between 1 and 4 Å, indicating the occupancy of N primarily among the N$_a$ sites instead of N$_d$ and N$_t$ sites (see **Figure 1a**), which results in similar Li-N distances compared to Li-O. The sharper Li-N peaks in LiPON compared to the Li-O peaks, around 2.1 and 3.74 Å (**Figure 1b**), indicate the presence of excess Li atoms near the N center, to compensate for the larger negative charge of the N$^{3-}$ compared to O$^{2-}$. Thus, our melt-quench AIMD simulations have successfully created snapshots of amorphous local environments which can be present in the actual LiPON phase, which should result in an accurate NequIP model.

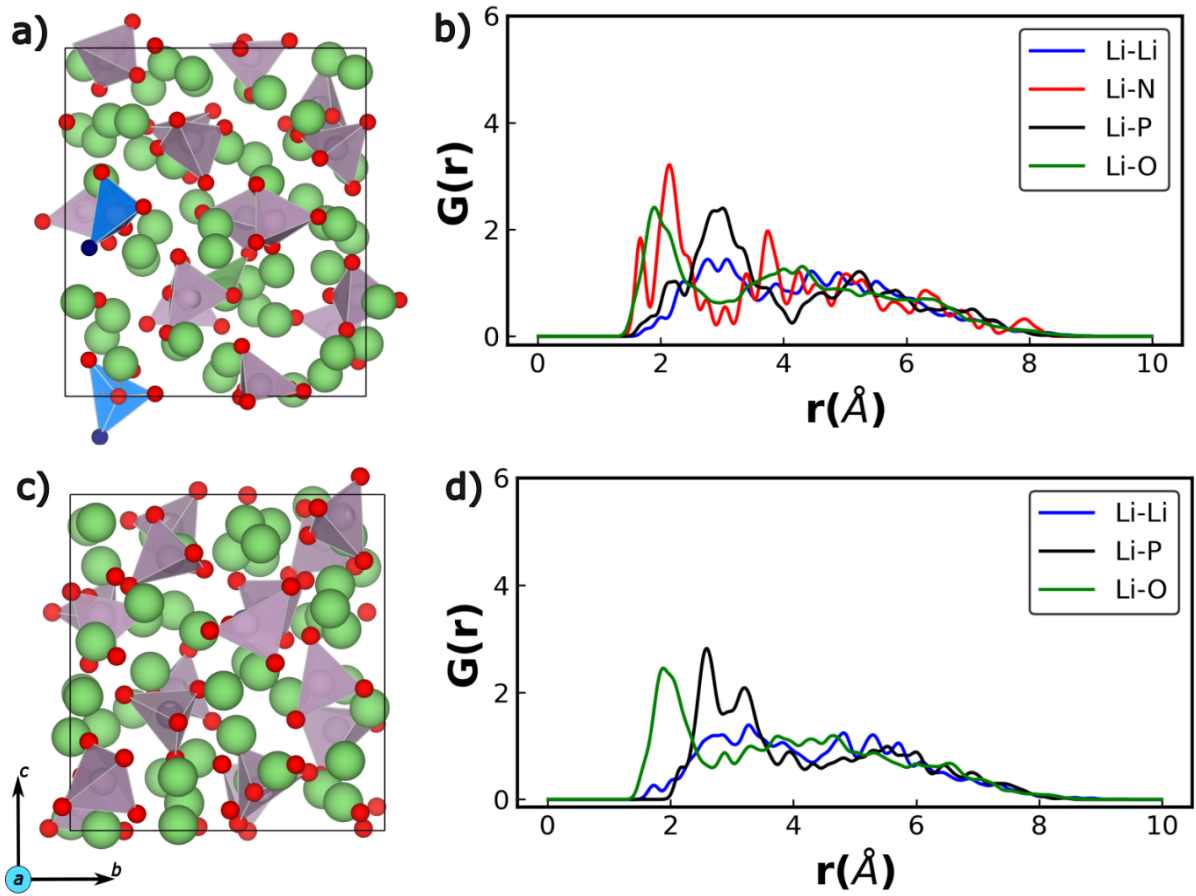

**Figure 1.** Sample AIMD-generated amorphous Li-rich LiPON structure (panel a) and amorphous Li$_3$PO$_4$ structure (panel c). Green, red, blue and violet spheres are Li, O, N, and P respectively. The PO$_4$ tetrahedra are in violet color with blue tetrahedra indicating the presence of N occupying the N$_a$ sites in P-based groups. The RDFs of LiPON and Li$_3$PO$_4$ are plotted in panels b and d respectively.

**Optimized Potential**

Using the set of optimized hyperparameters in **Table S2**, we trained two different NequIP models, one using equivariant tensor to generate the feature set during training (referred to as 'l_max = 2' within the NequIP architecture) and the other using invariant scalars (l_max = 0). Expectedly, our potential trained with l_max = 2 features displayed significantly lower training energy and force MAEs, of 5.5 meV/atom and 13.6 meV/Å, respectively, compared to the model with l_max = 0 features (energy and force MAEs of 22.3 meV/atom and 101.9 meV/Å). The lower training errors on the l_max = 2 model is attributed to the equivariant tensor features better capturing the local environments sampled by the model than the invariant scalars. Further, both the l_max = 2 and l_max = 0 models displayed consistently lower force errors on the validation set compared to the training set, with specific MAEs of 6.1 (16.1) meV/atom and 13.2 (95.4) meV/Å across energies and forces, respectively for the l_max = 2 (l_max = 0) model. The lower validation errors suggest that our models are likely not overfit on the training data. Given that the l_max = 2 model is more accurate than l_max = 0, we used the l_max = 2 for further analysis throughout the rest of the manuscript.

**Figure 2** displays the parity between the NequIP (optimized l_max = 2 model) predicted energies and DFT-calculated energies (panel a) and atomic forces (panel b) for the complete dataset (i.e., training and validation). The symbols in both plots indicate different subsets of the training set, namely, green squares for slabs, blue diamonds for the bulk stoichiometric phases, red triangles for the defective structures, orange circles for the AIMD simulations, and purple diamonds for the strained configurations. Overall, the NequIP-predicted energies are in strong agreement with DFT-calculations, with the exception of the defective (i.e., Li-rich/Li-poor) dataset. Similarly, the NequIP-predicted atomic forces are in good agreement with the DFT-calculated values, with the exceptions of the defective and strained datasets. The lack of agreement in predicted versus calculated forces in the strained subset is likely due to the large forces that are generated with the application of strain, with our NequIP model underestimating the calculated values. While the reason for disagreement between NequIP-predicted and DFT-calculated energies and forces within the defective dataset is unclear, we hypothesize that small perturbations in local bonding environments within these defective structures are resulting in larger variations in energies and forces, which is not fully captured by the NequIP model. Nevertheless, the overall training and validation errors exhibited by our NequIP model are close to those observed in literature,[35] across a wider and more diverse training set, suggesting that our model should be robust enough in modelling the amorphous LiPON energy landscape.

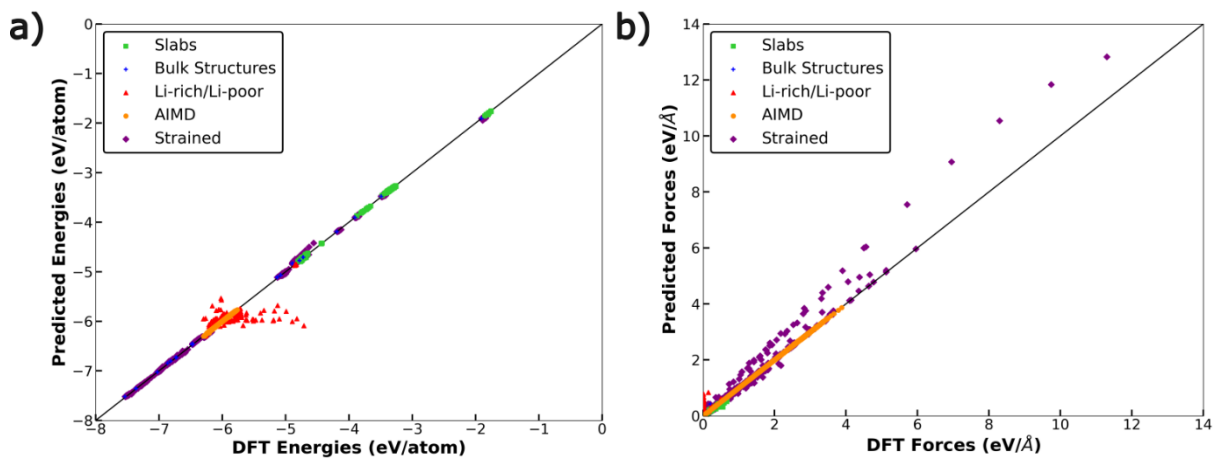

**Figure 2.** Parity between NequIP-predicted and DFT-calculated per-atom energies (panel a) and atomic forces (panel b) across the entire dataset. The different symbols indicate the different subsets of the dataset.

**Amorphous LiPON**

Using the optimized NequIP (l_max = 2) potential, we generate amorphous LiPON structures using the melt-quench MD approach. An example of a generated amorphous LiPON configuration (consisting of 124 atoms and a composition of $Li_{2.94}PO_{3.5}N_{0.31}$) that was melted at 2000 K and quenched to 250 K is displayed in **Figure 3a**, along with its RDF in **Figure 3b**. The notations used in **Figure 3** are similar to those used in **Figure 1**, with the violet, blue, dark green, and orange tetrahedra indicating $PO_4^{3-}$, $PO_3N^{4-}$

, $P_2O_7^{4-}$, and $P_2O_6N^{5-}$ groups, respectively. RDFs of structures melted to 600 K, 900 K, 1200 K, and 1500 K and quenched to 250 K are displayed in **Figure S2**, while RDFs of structures equilibrated at 600 K, 900 K, 1200 K, and 1500 K and subsequently quenched to 250 K are displayed in **Figure S3**.

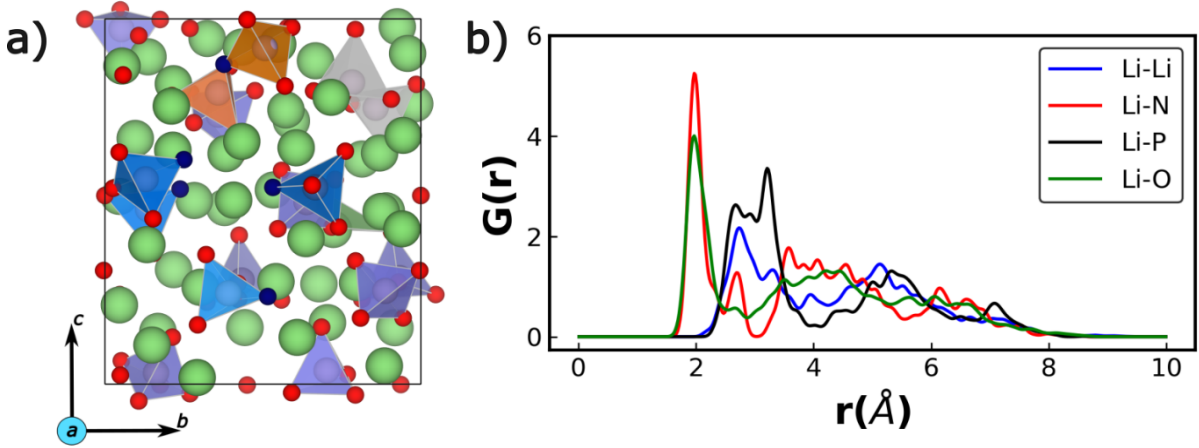

**Figure 3.** a) Example NequIP-generated amorphous LiPON structure at 250 K generated with a composition of $Li_{2.94}PO_{3.5}N_{0.31}$ via melt-quench MD simulations. Notations used are similar to Figure 1. Different P-based groups are highlighted with different colors: violet tetrahedra are $PO_4^{3-}$, blue tetrahedra are $PO_3N^{4-}$, gray polyhedra denote $P_2O_7^{4-}$ and orange polyhedra are $P_2O_6N^{5-}$ groups. b) The corresponding RDF of the generated LiPON structure.

From the structural snapshot in **Figure 3a**, we observe the nitrogen atoms to be incorporated within the LiPON framework mostly as $N_a$ (blue polyhedral), with few of the N sitting on $N_d$ sites bridging two phosphate groups together (orange polyhedra). This matches prior experimental analysis by Wang et al.[31] as well as the AIMD-based studies of Lacivita et al.[35,36] Moreover, we also observe phosphate tetrahedra linked by oxygen, resulting in the formation of $P_2O_7^{4-}$ groups (gray polyhedra), akin to the observations made by Marple et al.[37] Importantly, we do not observe any triply coordinated N sitting on $N_t$ sites in any of our melt-quench simulations, suggesting that it is unlikely to find N adopting this local coordination environment, which is in contrast to the initial experimental studies of Bates et al.[18] Nevertheless, the lack of N on $N_t$ sites is consistent with subsequent experimental and computational studies,[31,36,36,37] highlighting that our potential is generating reliable amorphous structures.

In terms of the RDFs (**Figure 3b**), we observe broad peaks beyond 4 Å for all types of neighbors to Li atoms, indicating the lack of long-range order, similar to our observation in AIMD simulations as well (**Figure 1**). Different from our AIMD simulations, we do observe sharper Li-O and Li-N peaks at ~2 Å, suggesting the formation of strong local order (or bonding) of Li atoms to nearby anions. Also, the Li-P RDF displays a broad shoulder and a peak towards ~2.9 Å, suggesting the existence of short-range order between the Li atoms and P-based groups. The differences between our AIMD and NequIP-based MD simulations can be primarily traced to the extent of equilibration done at the quenched temperature, with the MD simulations allowing the formation of local anionic clusters surrounding the Li atoms. With respect to quenching to higher temperatures (e.g., quenching to 600 K instead of 250 K), we observe a steady drop in the Li-O/Li-N peaks around 2 Å, while the broad shoulder observed for the Li-P at lower temperatures narrows into a sharper peak (**Figure S3**), suggesting a larger degree of short-range disorder when quenching to higher temperatures compared to 250 K.

**Li-diffusivity in amorphous LiPON**

We calculate the $Li^+$ diffusivity (and associated conductivity) in both the melt-quench and equilibrated LiPON configurations generated by the optimized NequIP model at 600 K, 900 K, 1200 K, and 1500 K, where the elevated temperatures allow a decrease in the simulation time required to capture Li-migration events. While the resultant natural logarithmic values of $D$ (in $cm^2/s$) and $\sigma$ (S/cm) values are plotted in **Figure 4** as a function of (inverse of) temperature. Solid (dashed) red and blue lines in **Figure 4** indicate the $D$ ($\sigma$) calculated in melt-quench and equilibrated structures, respectively. Given that $D$ and $\sigma$ are proportional to each other (see **Equations 1** and **2**), we observe similar trends in our calculated $D$ and $\sigma$.

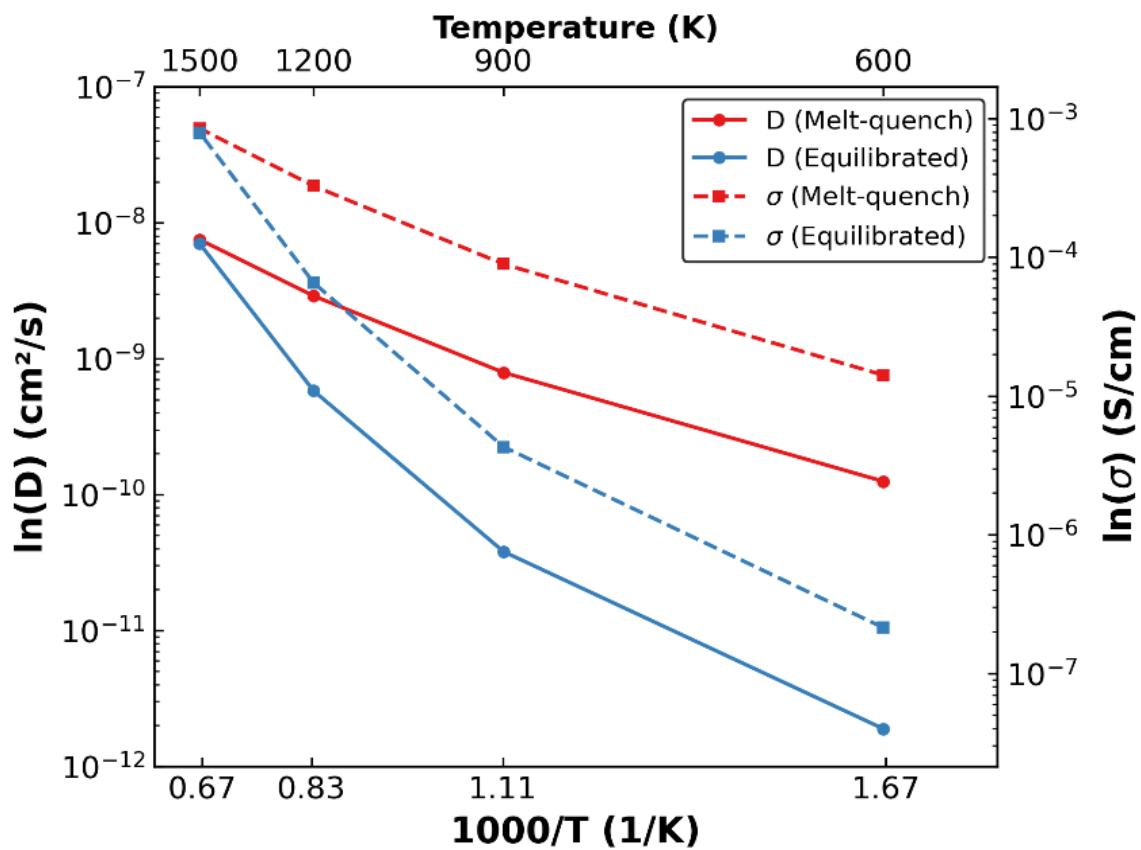

**Figure 4.** $Li^+$ diffusivity and ionic conductivity measured against temperature resulting from MD simulations conducted using the optimized potential.

Our calculated $D$ (and $\sigma$) values are consistently higher in the melt-quench structures (red lines in **Figure 4**) compared to the equilibrated structures (blue lines), except at the highest temperature simulated (i.e., 1500 K). For example, $D$ in melt-quench LiPON is ~two orders of magnitude higher ($1.25\times10^{-10}$ cm$^2$/s versus $1.89\times10^{-12}$ cm$^2$/s) than equilibrated LiPON. We can attribute the amorphous nature and the lack of significant long-range ordering (see **Figures S2 and S3**) to be a contributing factor to the higher $D$ (and $\sigma$) observed in the melt-quenched structures compared to equilibrated structures, which are in-line with experimental observations of superior Li-conductivity in amorphous-LiPON versus crystalline-Li$_3$PO$_4$.[27,59] In these structures, the amorphization leads to an orientational disorder in the phosphate tetrahedra destabilizing the Li ions.[19] The convergence of our calculated $D$ (and $\sigma$) values at 1500 K for both the melt-quench ($7.5\times10^{-9}$ cm$^2$/s) and equilibrated ($7.0\times10^{-9}$ cm$^2$/s) structures highlight that both structures become equally disordered at higher temperatures and exhibit similar local environments (see **Figures S2 and S3**), thus resulting in similar $D$ (and $\sigma$). The higher Li-diffusivity in LiPON compared to amorphous Li$_3$PO$_4$[60] can be attributed to the presence of N that facilitates Li migration. Finally, we observe our reported $D$ to be qualitatively similar to the $D$ values reported by Lacivita et al.[35] while our calculated $\sigma$ at 600 K marginally overestimates the experimental $\sigma$ at 298 K,[27] suggesting that our NequIP model is able to provide qualitatively accurate trends with reasonable quantitative accuracy.

**Interface model and Li-diffusivity**

**Figure 5a** shows the generated Li(110)||LiPON interface, with bulk Li metal constituting the 'left' portion of the structure and bulk LiPON (generated via melt-quench) constituting the 'right' portion. Green, purple, red, and blue spheres are Li, P, O, and N, respectively. Panels b and c of **Figure 5** plot the average Li-Li bond length and the average number of Li neighbors to a given Li atom across the structure. Note that we average the bond length and number of Li neighbors at a given $c$-axis value (i.e., averaged across an $a$-$b$ plane), where the $c$-axis is perpendicular to the interface. The pink, blue, and yellow shades in **Figures 5b** and **c** indicate regions of bulk Li, bulk LiPON, and the interface (i.e., transition from Li to LiPON), respectively.

Importantly, we observe the average Li-Li bond distances and average number of Li neighbors to be uniform within the bulk Li region (shaded pink regions), highlighting the crystalline nature of body centered cubic Li metal. The decrease (increase) in bond distances (number of neighbors) in the bulk Li region close to zero value of the $c$-axis is due to periodic boundary conditions utilized in our calculations. Across the transition region (shaded yellow regions), we observe sharp changes in both Li-distances and neighbors. Further, both bond distances and Li neighbors exhibit non-monotonic trends within bulk LiPON (blue shaded regions), attributable to the amorphous nature of the structure.

The NequIP-calculated $D$ of the Li(110)‖LiPON interface is plotted in **Figure 6**, where we distinguish between $D$ in the bulk phases (i.e., bulk Li and bulk LiPON, red lines) and across the transition region (or interface, blue lines) between Li and LiPON. Specifically, we calculate the $D$ along the $c$-direction (i.e., perpendicular to the interface) and across the transition region to quantify Li transport from bulk Li to bulk LiPON (and vice-versa), which in turn should correspond to how conductive to Li is the Li(110)‖LiPON interface. Thus, the calculated $D$ in the bulk phases correspond to Li migrations along $a − b$ planes (i.e., parallel to the interface). The variation of the MSD of Li with $\Delta t$ at 300 K, 600 K, and 900 K are compiled in **Figure S4**.

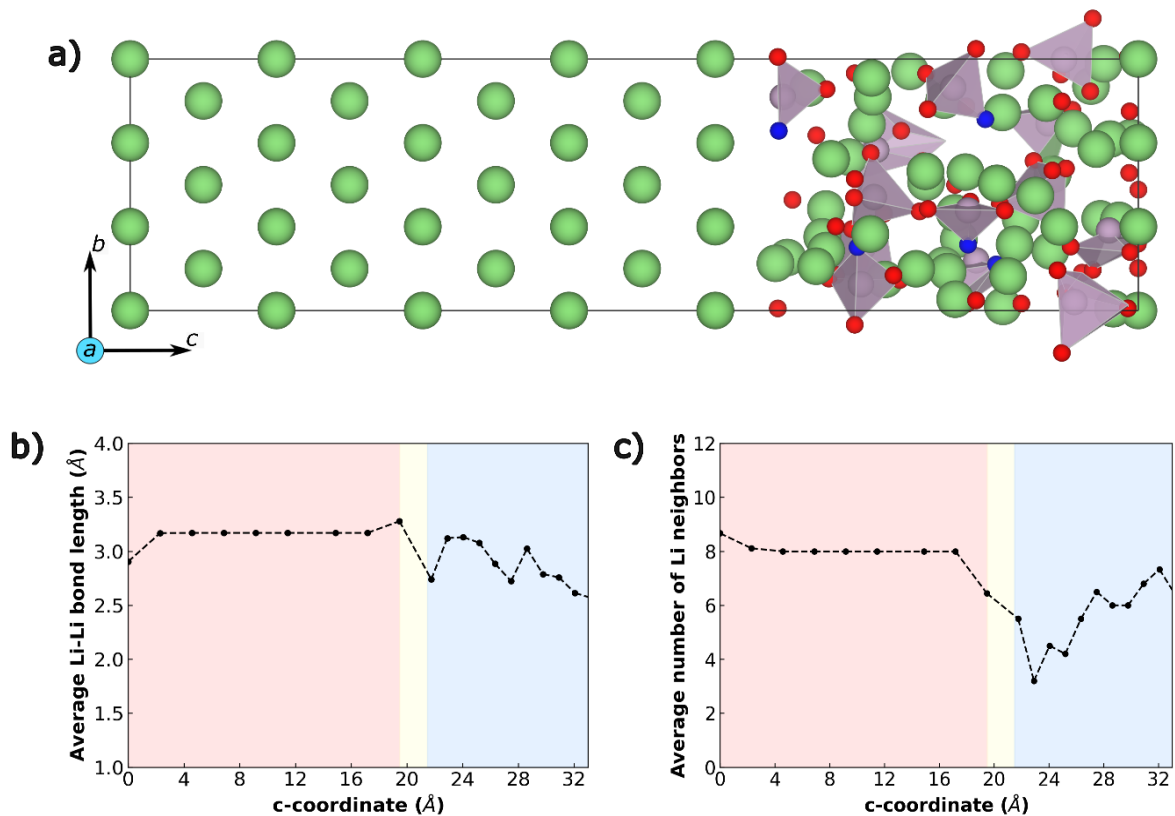

**Figure 5**. a) The Li-(110)‖LiPON interface. b) Variation in the average Li-Li bond length and c) variation in the average number of Li neighbors along the $c$-axis.

Importantly, we find $D$ in the bulk phases to be higher than the transition region (~$10^{-8}$ cm$^2$/s versus $10^{-9}$ cm$^2$/s), which is expected given Li is known to diffuse reasonably well in its bulk metallic state and within amorphous LiPON. Across the transition region, the drop in $D$ that we observe is marginal (i.e., one order of magnitude) and should not significantly affect the transport of Li from one bulk phase to another. However, our calculated $D$ are lower than reported values across the Li(100)‖Li$_3$P interface (~$10^{-6}$ to $10^{-7}$ cm$^2$/s) and are similar in magnitude to Li(110)‖Li$_2$S and Li(110)‖LiCl interfaces (~$10^{-7}$ to $10^{-9}$ cm$^2$/s).[61] Given that Li-interfaces with Li$_2$S and LiCl are known to be poor for Li-transport, resulting in impedance increase in argyrodite-type electrolytes,[61] we expect the Li‖LiPON interfaces to also be active only under low-rate conditions, in agreement with the usage of LiPON in thin-film devices.[62] Thus, further changes in composition and/or preconditioning of the Li||LiPON interface may be deployed for improving the power performance of LiPON electrolytes in practical devices.

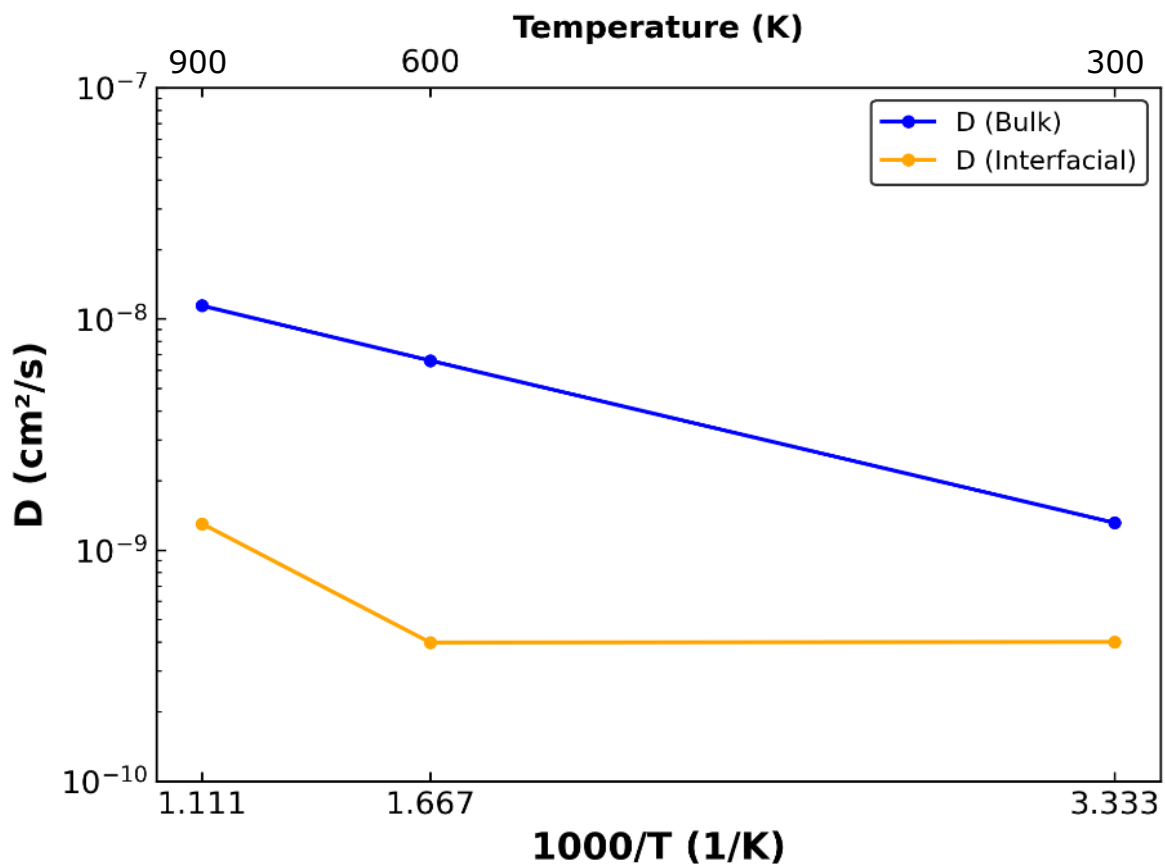

**Figure 6**. Variation of Li$^+$ diffusivity through bulk phases and across the interfacial (transition) region of the Li(110)||LiPON interface calculated at 300K, 600K and 900K.

## Discussion

LiPON's excellent electrochemical properties when compared to crystalline-Li$_3$PO$_4$, including its ability to suppress cycle life degradation and dendrite formation over thousands of charge-discharge cycles has made it a highly appealing solid electrolyte in thin film devices. In this study, we have created a DFT-calculated dataset of over 13,000 configurations which span the different local environments that may be encountered in an amorphous LiPON structure. We used the DFT-dataset to train an equivariant NequIP potential and in turn used the NequIP model to describe Li-transport in bulk amorphous LiPON and across a Li(110)||LiPON interface. Specifically, we generated bulk amorphous LiPON structures of composition Li$_{2.94}$PO$_{3.5}$N$_{0.31}$ using melt-quench MD simulations and also via an equilibration approach which retained a higher degree of long-range order. Importantly, we found N to occupy the N$_a$ and N$_d$ sites only, and did not observe any occupation of N$_t$ sites. Further, our calculated Li-diffusivities (and conductivities) in bulk LiPON are in qualitative agreement with previous studies while we observe Li-diffusivity across a Li(110)||LiPON interface to be ~one order of magnitude slower than the bulk phases. We hope our work highlights the high accuracy, transferability, and efficiency of equivariant GNN-based MLIPs and motivates utilization of these architectures in studying other complex amorphous structures.

The choice of dataset plays a crucial role in the training and validation of any MLIP model. It is important for the dataset to contain a diverse set of chemically relevant configurations and local bonding environments, enabling the MLIP model to learn the PES effectively. In this regard, we observe NequIP to be highly data-efficient owing to its equivariant GNN-based architecture which allows for features to be propagated beyond the chosen cutoff-radius via message passing. With a dataset of just over 13,000 DFT-generated configurations, the NequIP models not only achieve training energy and force errors (MAE) similar to what is typical of DFT, but also show remarkable resistance to overfitting. Additionally, our DFT-based dataset could be used to train other MLIPs such as CHGNet[63], Allegro[64], DimeNET[65] and MACE[66] to benchmark the performance of different GNN-based architectures. In any case, there do exist certain gaps in our training dataset, which when supplemented with additional data could lead to even better potentials. For example, prior studies have indicated the Li-LiPON interface to be passivized by the presence of decomposition products such as $Li_2O$, $Li_3N$, $Li_3P$ and $Li_3PO_4$.[67] While AIMD-based configurations and individual slabs of these systems are part of our training dataset, their interfaces with pure Li or LiPON, which are computationally expensive to compute are lacking. Thus, expansion of the dataset used in this work will result in better insights and more accurate predictions of the kinetics of the Li-LiPON system.

Along with the dataset limitations, improvements in hyperparameter tuning could further improve the accuracy and usage of the trained potentials. One bottleneck in hyperparameter tuning, and the general process of training MLIPs, is the computational cost of training. For instance, though NequIP's potentials are highly data-efficient due to their use of equivariant tensors, their utilization of equivariant tensor-based features leads to an exponential increase in the required training time. Training on one NVIDIA V100 Tesla 16 GB GPU card (with 192 GB of RAM and no hyperthreading) we observed training sessions to take as long as 36 hours with tensor features, while using only invariant scalar features typically takes only a few minutes of training. However, we have shown that NequIPs with scalar features have significantly higher training and validation errors compared to NequIPs with tensor features. Hence, there exists a need for computationally optimizing the MLIP architectures such that the models can be trained quickly.

Experimentally, $Li^+$ diffusivity is significantly affected by the percentage of crystallinity, experimental conditions, and the quality of the LiPON samples. Notably, the composition of LiPON (ratio of anions to P, Li content, and number of isolated O), lattice disorder, and defects are known to play a crucial role in Li$^+$ diffusivity.[35] Considering the induced disorder in the phosphate tetrahedra of LiPON due to the presence of $N_a$ and $N_d$ sites, it is possible that a similar phenomenon is at play in the melt-quench LiPON structures, which leads to its higher diffusivity compared to the equilibrated LiPON structures in our work. The presence of $N_d$ sites could also contribute to this variation in diffusivity, as the equilibrated structures contain only $N_a$ sites, compared to the melt-quench configurations that contain both $N_a$ and $N_d$ sites. This variation in how nitrogen is incorporated into different LiPON structures comes from the trained potential itself and the temperatures exposed during the MD runs. Nonetheless, the presence of $N_d$ sites in the LiPON framework has been known to promote Li$^+$ diffusivity[35] attributed to Li$^+$ being less tightly bound in the vicinity of $N_d$ sites compared to $N_a$ or O sites.

## Conclusion

Amorphous LiPON is an important class of Li solid electrolytes that are known to provide reasonable cycle life and power performance in thin film devices while suppressing the growth of Li dendrites. In this work, we developed GNN-based NequIPs to investigate Li$^+$ transport in bulk amorphous LiPON electrolytes and across a Li(110)||LiPON interface. We generated a DFT-based dataset consisting of 13,454 structures and randomly split the dataset into 90:10 training:validation sets to train and optimise the NequIPs. Note that our training dataset comprised of bulk and strained structures, Li-rich and Li-poor defective structures, LiPON-like amorphous configurations from AIMD, and slabs. Importantly, we observed the trained NequIPs, with equivariant tensor features, to be highly accurate with training (validation) energy and force MAEs of 5.5 (6.1) meV/atom and 13.6 (13.2) meV/ Å, respectively. Subsequently, we used the trained NequIPs to generate amorphous LiPON structures that exhibited N occupation of $N_a$ and $N_d$ sites, consistent with prior AIMD and experimental results. Our Li-diffusivity estimates in bulk LiPON were qualitatively similar to previous studies as well, and we observed Li-diffusivity to improve with increasing disorder in the LiPON structure. Further, we modelled Li$^+$ transport across the Li(110)||LiPON interface, where we noticed Li-movement across Li to LiPON (or vice-versa) to be one order of magnitude slower than bulk motion within metallic Li and bulk LiPON. Thus, we do not expect the Li||LiPON interface to be insulating towards Li motion and any associated impedance buildup should be negligible. Finally, our study demonstrates that MLIPs can act as promising tools to model amorphous solid electrolytes as well as their associated interfacial behavior and we hope our work instigates further studies exploring amorphous materials.

## Acknowledgements

G.S.G acknowledges financial support from the Indian Institute of Science (IISc) and support from the Science and Engineering Research Board (SERB) of the Government of India, under Sanction Numbers

SRG/2021/000201 and IPA/2021/000007. All the density functional theory calculations used in generating the dataset were performed with the computational resources provided by Supercomputer Education and Research Center, Indian Institute of Science. The authors gratefully acknowledge the computing time provided to them on the high-performance computers noctua1 and noctua2 at the NHR Center PC2. This was funded by the Federal Ministry of Education and Research and the state governments participating on the basis of the resolutions of the GWK for national high-performance computing at universities (www.nhrverein.de/unsere-partner). The computations for this research project were performed using computing resources under project, hpc-prf-emdft.

## Data Availability

All calculated data files and trained potentials are available to the public freely via our GitHub repository: https://github.com/sai-mat-group/ann-lipon.